\documentclass[12pt,a4paper]{article}
\usepackage{amsmath,amssymb}
\usepackage{epsfig,graphicx}
\makeatletter
\newcommand*{\rom}[1]{\expandafter\@slowromancap\romannumeral #1@}
\makeatother

\begin{document}
\title{\bf Magnetic Field Induced Effects in Quark Matter\footnote{Contribution to the Proceedings of ``QUARKS 2012'' International Seminar, Yaroslavl, June 4-10, 2012}}
\author{B.~O.~Kerbikov$^{a,b}$\footnote{{\bf e-mail}: borisk@itep.ru},
  M.~A.~Andreichikov$^{a,b}$\footnote{{\bf e-mail}:
    andreichicov@mail.ru}
  \\
  $^a$ \small{\em State Research Center Institute for Theoretical and Experimental Physics} \\
  \small{\em Bolshaya Cheremushkinskaya, 25, 117218 Moscow, Russia }
  \\
  $^b$ \small{\em Moscow Institute of Physics and Technology} \\
  \small{\em Institutskii per., 9, Dolgoprudny, 141700 Moscow Region,
    Russia}}

\date{}

\maketitle

\begin{abstract}
  Quark--gluon matter produced in relativistic heavy--ion collisions(RHIC and
  LHC) is subject to a super--strong magnetic field(MF) $\sim
  10^{18} - 10^{20}\ G$. Quark matter(QM) response to MF allows to get
  a new insight on its properties. We give a cursory glance on MF
  induced effects.
\end{abstract}

Relativistic heavy--ion collisions(RHIC and LHC) generate gigantic
magnetic field (MF) $eB \sim \Lambda^2_{QCD}$.  The response of QM
to such a field has been intensively studied during the last several
years. In this brief presentation we merely give a list of the
corresponding problems. In no way this material can be considered as a
review paper. We apologize for the absence of references. A list of
$10^2$ references would be excessive for this format.

We begin by comparing MF-s encountered in Nature and in the
Laboratory.

\begin{center}
  The Hierarchy of MF-s (in Gauss)
\end{center}

\begin{table}[h]
  \centering
  \begin{tabular}{l l}
    Medical MPI scan & $10^4$ \\
    ATLAS at LHC & $4 \cdot 10^4$ \\
    Lab.(preserving the equipment) & $10^6$ \\
    Lab.(explosion) & $28 \cdot 10^6$ \\
    Schwinger(for electron) & $4.4 \cdot 10^{13}$ \\
    Surface of magnetars & $ 10^{14}-10^{15}$ \\
    RHIC and LHC & $10^{18}-10^{20}$ \\
    Early Universe & $10^{24}$ \\
  \end{tabular}
\end{table}

From the above table we see that quark--gluon matter produced in
heavy--ion collisions is embedded in the strongest possible MF. This
field lasts for only $\tau \sim 0.2 \ fm$ unless the conductivity of
the produced matter is high enough(see below). It turns out that
effects caused by MF at $\tau \le 0.2 \ fm$ almost evaded a thorough
investigation. It might be due to the fact that at this stage of the
``fireball'' evolution its nature and dynamics are rather
complicated. In fact, two drastic oversimplifications have been done
in most studies: 1) MF was assumed to be constant both in time and in
space direction, and 2) an infinite system in thermodynamic
equilibrium has been usually considered. Probably the rare exception
is the Chiral Magnetic Effect(CME) which implies the creation of
flashing topological charges in hot and dense medium. However,
microscopic derivation of topological effects is lacking, while the
lattice calculations of CME were performed for low temperature and in
thermodynamic equilibrium(see Section 4 below).

Now we start the list of MF induced effects.

\section{MF decay rate}

MF duration $\tau \sim 0.2 \ fm$ corresponds to the time of maximal
overlap of the colliding nuclei. Magnetic response of the produced
matter can make this time an order of magnitude larger. On the
dimensional grounds the decay time of MF is $\tau' \simeq \sigma
L^2$, where $\sigma$ is the electrical conductivity, $L$ is the
characteristic length scale of the spatial variation of MF. If for a
rough estimate we take $\sigma/T \simeq 0.5,\ T \simeq 200 \ MeV,\ L
\simeq 2 \ fm$, we obtain $\sigma' \simeq 2 \ fm$. In presence of
magnetic monopoles(possibly seen on the lattice) this time will be
shorter.

\section{Phase space arguments}

MF stronger than Schwinger critical field ($B_c = m_e^2/e =
4.4 \cdot 10^{13} \ G $ for electron) results in enlarging of the phase
space available for the electron in $\beta$-decay and in the
corresponding increase of the decay rate. This is due to the Landau
orbits phase space. The same effect for quark emerging from a decay
has not been investigated.

Another phenomenon concerns the population of Landau levels in dense
QM. The dispersion relation for quark in MF reads

\begin{equation}
  \label{eq:1}
   \omega_{n,\sigma}(k_z) = [k_z^2 + m_f^2 + q_fB(2n + 1 + \sigma)]^{1/2},
\end{equation}
where $ \mathbf{B} || \mathbf{z}$, $f$ is the
flavour index, $q_f$ is the absolute value of quark electric charge,
$\sigma = \pm 1$. Consider a dense QM at low temperature, i.e., in the
regime $\mu \gg T$, where $\mu$ is the chemical potential. Condensed
matter wisdom tells that the key physical processes(like transport)
are determined by the vicinity of the Fermi surface. If in (1) we take
$\omega_{n \sigma} = \mu_F = \sqrt{k_F^2 + m_f^2}$, then only Landau
levels up to
\begin{equation}
  \label{eq:2}
   n_{max} = \frac{\mu_F^2 - m_f^2}{2q_fB}
\end{equation}
survive. For example,
for $\mu_F = 3 m_{\pi},\ q_fB = 5m_{\pi}^2,\ m_f = 0$, one gets
$n_{max} = 1$. We note that the dominance of the Lower Landau
Lavel(LLL) in strong MF has a general nature not inferred from
(2). Another related general feature of strong MF in the transverse
shrinkage of the system and dimensional reduction $3d \rightarrow 1d$.

\section{QCD phase diagram in MF}

Before going to concrete results on QCD phase diagram, a remark has to
be done. With rare exceptions, QCD phase diagram has been studied
under a tacit assumption that the system is infinite and in a state of
thermodynamic equilibrium. Such an approach might have been
appropriate for neutron stars, but reliable lattice calculations are
only possible for $\mu = 0$. For $\mu > 0$ one has to resort to
models, like NJL, and no clear cut conclusions are available.

The first point is the influence of MF on light quark condensate
$\langle \bar \psi \psi \rangle$. Until recently it seemed firmly
established that $\langle \bar \psi \psi \rangle$ is increasing with
$B$. Correspondingly the critical temperature $T_{\chi}(B)$ also
grows. This kind of response got the name of ``magnetic
catalysis''. The new lattice calculations revealed more complicated
picture. Magnetic catalysis was confirmed at low temperature, while
around $T_{\chi}$ the $B$-dependence of $\langle \bar \psi \psi
\rangle$ is not monotonous resulting in the decrease of
$T_{\chi}$. This might be due to indirect interaction between gluons
and MF. As already mentioned, at $\mu > 0$ only the results of model
calculations are available. There is an indication that due to MF the
first--order transition line, which starts at the critical point, goes
up.

At $\mu = 0$ the phase transition is an analytic crossover. The chiral
$T_{\chi}$ and deconfinement $T_L$ temperatures are splitted, most
studies show that $T_L > T_{\chi}$. The conclusions of different
authors on the MF dependence of $T_L$ are contradictory. The latest
lattice calculation indicates a reduction of $T_L$ in MF.

Next we shall consider several more specific problems.

\section{Chiral magnetic effect(CME)}

It certainly provoked a record wave of discussions among all MF
induced effects. It also brought to light the fact that heavy--ion
collisions generate a super--intense MF. It still needs a sound
experimental confirmation and work in this direction is in
progress. In the most concise form CME is represented by the formula
\begin{equation}
  \label{eq:3}
   \mathbf{j} = N_c \sum_f \frac{q_f^2\mu_5}{2\pi^2} \mathbf{B}, 
\end{equation}
where $\mathbf{j}$ is the electric current, $\mu_5$ is the chiral
chemical potential which induces a difference in number between
right--handed and left--handed particles. On the theoretical side
equation (3), or similar ones, were obtained starting from different
basic ideas: topological charge, axial anomaly, Chern-Simons action,
strong $\theta$--angle, etc. It also turned out that equations like
(3) were discovered much earlier. As we already mentioned, lattice
calculations can hardly be considered as a direct evidence of CME
since simulations of chiral fermions at high temperature is out of
reach for present lattice calculations.

\section{Conductivity in MF}

In Section 1 it was shown that the decay of MF depends on the value of
the electrical conductivity(EC). Latice calculations at $\mu=0$ and
$T$ around the phase transition temperature or somewhat higher give
$\sigma/T \simeq 0.3-0.4$ which corresponds to MF decay time of a few
fm. Another lattice group calculated EC in the same $(T,\ \mu)$ region
with MF. They obtained much less value for EC and very weak dependence
on MF. The last fact didn't get a physical explanation. It has a
natural explanation in a different regime described in Section 2,
namely high density and low temperature. Here the EC can be decomposed
into two contributions: the Drude and the quantum ones. Drude part is
calculated using Kubo formula and MF dependence enters via the
combination $(eB/\mu)^2 \tau^2$, where $\tau$ is the momentum
relaxation time. As a result, MF dependence becomes significant only
at $eB \ge 5 m_{\pi}^2$. The quantum part depends on the MF via
$1/l_B$, $l_B = (eB)^{-1/2}$, so that it has a relatively weak square
root dependence on MF. We note that quantum contribution may be
negative. In high density regime quantum EC is dominated by
fluctuating (precursor) Cooper pairs. The same mechanism is
responsible for another spectacular effect which we consider in the
next section.

\section{Giant Nernst Effect}

Consider $\mathbf{B} || \mathbf{z}$ and the temperature gradient
$\nabla_xT$. Then counterpart of Hall effect is the
Nernst--Ettingshausen one. It amounts to the induction of the
electric field $E_y$ and is characterized by a coefficient
\begin{equation}
  \label{eq:4}
   \nu = \frac{E_y}{(-\nabla_xT)B}
\end{equation}
It was shown by Varlamov and
co-authors that fluctuating pairs lead to a giant effect. In heavy ion collisions the electric field will influence the particle spectra.  The
corresponding work is in progress.

The effects listed in sections 1--6 do not cover the whole subject. In
particular, left in the cold are:

\begin{enumerate}
\item Magnetic tuning of BCS-BEC crossover
\item Quarkonium dissociation via ionization in MF.
\item Enhancement of flaw anisotropies due to MF.
\item QM viscosity in MF.
\item ...

\end{enumerate}

B.K. is grateful to the remarks and criticism received when the
material was presented at the \rom{46} PNPI Winter School,
``QUARK--2012'' seminar, and at the seminars in ITEP, BNL, Stony
Brook.

\end{document}